\def \no{\noindent} 
 
\def \yskip{\penalty-50\vskip3pt plus 3pt minus 2pt}
\def \reference{\par \yskip \noindent \hangindent .4in \hangafter 1}
\def \abc#1#2#3#4 {\reference#1, {\sl#2}, {\bf#3}, #4}
\def \blank {\lower 5pt\hbox to 0.75in{\hrulefill}}
\def \kms{~\rm{km}~\rm{s}^{-1}}

\def \cm{~\rm{cm}}
\def \s{~\rm{s}}
\def \km{~\rm{km}}

\def \g{~\rm{g}}
\def \AU{~\rm{AU}}
\def \yrs{~\rm{yrs}}
\def \yr{~\rm{yr}}
\def \K{~\rm{K}}

\def \lae{\mathrel{<\kern-1.0em\lower0.9ex\hbox{$\sim$}}}
\def \gae{\mathrel{>\kern-1.0em\lower0.9ex\hbox{$\sim$}}}

\documentstyle[11pt,aasms,tighten,flushrt]{article}
\begin{document}
%\normalsize
\small

\setcounter{page}{1}
%\noindent Presented at the 180 IAU Symposium: {\it Planetary Nebulae}, 
%August 1996.
%dust1.tex
\begin{center} \bf 
STELLAR STRUCTURE AND MASS LOSS \\
ON THE UPPER ASYMPTOTIC GIANT BRANCH 
\end{center}
%\vspace*{2.0cm}

\begin{center}
 Noam Soker and Amos Harpaz \\
Department of Physics, University of Haifa at Oranim\\
%Mathematics-Physics\\
Oranim, Tivon 36006, ISRAEL \\
soker@physics.technion.ac.il 
\end{center}

%\clearpage 

\begin{center}
\bf ABSTRACT
\end{center}

 We examine the envelope properties of asymptotic giant branch 
(AGB) stars as they evolve on the upper AGB and during the
early post-AGB phase.
 Due to the high mass loss rate, the envelope mass decreases by more
than an order of magnitude.
 This makes the density profile below the photosphere much shallower,
and the entropy profile much steeper.
 We discuss the possible role of these changes in the profiles in
the onset of the high mass loss rate (superwind) and the large 
deviation from spherical mass loss at the termination of the AGB.
 We concentrate on the idea that the shallower density profile and steeper
entropy profile allow the formation of cool magnetic spots, above
which dust forms much more easily.

\noindent
%{\it Subject heading:}   % to APJ

%\clearpage 

% ======================================================================
\section{INTRODUCTION}
% ======================================================================

 The inner regions of many planetary nebulae (PNs) and proto-PNs
show much larger deviation from sphericity than the outer regions
(for recent catalogs of PNs and further references see, e.g., Acker {\it et al.} 1992;
Schwarz, Corradi, \& Melnick 1992; Manchado {\it et al.} 1996;
Sahai \& Trauger 1998; Hua, Dopita, \& Martinis 1998).
 By ``inner regions'' we refer here to the shell that was formed from
the superwind$-$the intense mass loss episode at the termination of the
AGB, and not to the {\it rim} that was formed by the interaction with the
fast wind blown by the central star during the PN phase
(for more on these morphological terms see Frank, Balick, \& Riley 1990).
 This type of structure, which is observed in many elliptical PNs,
suggests that there exists a correlation, albeit not perfect, between the
onset of the superwind and the onset of a more asymmetrical wind.
  In extreme cases the inner region is elliptical while the outer
region (outer shell or halo) is spherical (e.g., NGC 6826, Balick 1987).
 Another indication of this correlation comes from spherical PNs. 
 Of the 18 spherical PNs listed by Soker (1997, table 2),
$\sim 75 \%$ do not have superwind but just an extended spherical halo.

 The transition to both high mass loss rate and highly non-spherical
mass loss geometry may result from the interaction with a stellar
or substellar companion (Soker 1995; 1997), or
from an internal property of the AGB star.
   Regarding the second possibility, the change in the mass loss properties
has been attributed to the decrease in the envelope density
(a recent different suggestion for this behavior made by Garcia-Segura
{\it et al.} [1999] is criticized in $\S 3.2$ below).
  In an earlier paper (Soker \& Harpaz 1992) we proposed a 
mode-switch to nonradial oscillations.
 This scenario is related to the fast
decrease in the Kelvin-Helmholtz time of the envelope, eventually becoming
shorter than the fundamental pulsation period.
 For a low mass envelope the Kelvin-Helmholtz time is 
\begin{equation}
\tau_{\rm KH} ({\rm envelope}) \simeq
\frac{G M_{\rm core} M_{\rm env} } {R L }
= 0.6
\left( \frac {M_{\rm core}}{0.6 M_\odot} \right)
\left( \frac {M_{\rm env}}{0.1 M_\odot} \right)
\left( \frac {R}{300 r_\odot } \right)^{-1}
\left( \frac {L}{10^4 L_\odot } \right)^{-1}
\yrs,
\end{equation}
where $M_{\rm core}$ and $M_{\rm env}$ are the core and envelope mass,
respectively, $R$ the stellar radius and $L$ stellar luminosity.
 As the envelope expands the fundamental  mode period increases, and due
to mass loss and increase in luminosity, the Kelvin-Helmholtz time
decreases.
 When the envelope mass becomes $\sim 0.1 M_\odot$, the Kelvin-Helmholtz
time becomes shorter than the fundamental pulsation period.
  We noted (Soker \& Harpaz 1992), though, that even this
``single-star'' mechanism requires a
binary companion to spin-up the AGB envelope in order to fix the symmetry
axis, and make some non-radial modes more likely than others
(if all non-radial modes exist, then the overall mass loss geometry
is still spherical). 

 {{{   The widely accepted model for the high mass loss rate on the upper
AGB includes strong stellar pulsations coupled with efficient 
dust formation (e.g., Wood 1979;  
Bedijn 1988;  Bowen 1988; Bowen \& Willson 1991;
Fleischer, Gauger, \& Sedlmayr 1992; H\"ofner \& Dorfi 1997). }}}
   In a recent paper, one of us (Soker 1998) proposed another mechanism
{{{  for asymmetrical mass loss }}} based on the decrease of
the envelope density.
{{{ This mechanism is based on the above model of mass loss due to
pulsations coupled with dust formation. }}}
 In that {{{  scenario }}}
which was further developed by Soker \& Clayton (1999),
Soker assumes that a weak magnetic field forms cool stellar spots, which
facilitate the formation of dust closer to the stellar surface.
 Dust formation above cool spots, due to large convective elements
(Schwarzschild 1975), or magnetic activity, enhances the mass loss
rate (Frank 1995).
 If spots due to the dynamo activity are formed mainly near the equatorial
plane (Soker 1998; Soker \& Clayton 1999), then the degree of deviation
from sphericity increases.
 Soker (1998) claims, {{{  based on a crude estimate, }}}
that this mechanism, of dust formation above
cool magnetic spots, operates for slowly rotating AGB stars,
having angular velocities of $\omega \gtrsim 10^{-4} \omega _{\rm Kep}$,
where $\omega_{\rm Kep}$ is the equatorial Keplerian angular velocity.
 Such angular velocities could be gained from a planet companion of
mass $\gae 0.1 M_{\rm Jupiter}$, which deposits its orbital angular
momentum to the envelope at late stages, or even from single stars 
which are fast rotators on the main sequence.   

 The advantage of the late tidal interaction with a binary companion
is that the companion can explain the formation of jets
(Soker 1992, 1997; Soker \& Livio 1994), while in models based on the
decrease in the envelope mass (where a companion is required only
to spin-up the envelope at earlier stages), there is no satisfactory
model for the formation of jets.
 In any case, even for a late tidal interaction the mechanism of dust
formation above cool magnetic spots may be significant.

{{{   The scenario of asymmetrical mass loss via cool magnetic spots
contains several assumptions and speculative effects, e.g., about
the dynamo activity.
 To strengthen this scenario, we conduct numerical calculations to study
the properties of the envelopes of upper AGB stars.
 We find that the entropy gradient becomes
steeper, while the density profile becomes shallower.
 The steeper entropy profile increases the convective pressure,
and makes the envelope, inside and outside the magnetic flux tubes,
more prone to any convective motion.
  We speculate that these changes will result in a more efficient
amplification of the envelope magnetic field, both through global dynamo
activity and local concentration of magnetic flux tubes by the
convective motion. }}}
 The evolutionary simulation is described in $\S 2$, where
we follow the evolution of an AGB star,
from the late AGB to the post-AGB phase.
 We present the structure of the envelope at five evolutionary points,
and show that the envelope's density and entropy profiles changed
significantly, and therefore may be the properties which determine the
large change in the behavior of the mass loss rate and geometry at the
end of the AGB.
 We would like to stress that we do not propose a new mass loss mechanism.
 We accept that {{{  pulsations coupled with }}} radiation
pressure on dust is the mechanism for mass loss
{{{  (e.g., Bedijn 1988;  Bowen 1988),}}}
and that the luminosity, radius, and mass of the AGB star are the main
factors which determine the mass loss rate
{{{  (e.g.,  Bowen \& Willson 1991; H\"ofner \& Dorfi 1997). }}}
 We only suggest that magnetic cool spots ease the formation of dust,
and that their concentration near the equator causes the mass loss
geometry to deviate from sphericity (Soker 1998; Soker \& Clayton 1999).
  In $\S 3$ we discuss the general behavior of the envelope properties,
in particular the density and entropy profiles, and
{{{  speculate on }}} the way by which they
may enhance the formation of cool magnetic spots.
 We summarize in $\S 4$.

% ======================================================================
% ======================================================================
\section{ENVELOPE PROPERTIES}
% ======================================================================

 In this section we describe the results of a numerical simulation
of an AGB stellar model, as it evolves on the upper AGB, and turns to
a post-AGB star.
 The stellar evolutionary code is similar to the one described by 
Harpaz, Kovetz, \& Shaviv (1987), and we describe it here briefly. 
The equations of evolution are replaced by the difference equations for 
each mass shell:   
\begin{equation}
 T(S - S^0) = \left[ q - {\Delta L \over \Delta m} + 
 T\sum_a \left( {\partial S\over \partial x_a} R_a \right) \right] \delta t,   
\end{equation}
\begin{equation}
(v - v^0) =
\left[ -4\pi r^2 {\Delta P\over \Delta m} - {G m \over r^2} \right]
 \delta t, 
\end{equation}
\begin{equation}
{\Delta(4\pi r^3/3)\over \Delta m} = {1\over \rho},   
\end{equation}
and
\begin{equation}
(x_a - x_a^0) = [R_a(\rho^0, T^0, x^0)] \delta t,     
\end{equation}
where $\delta t$ is the selected time step, $\Delta$ denotes space 
difference, superfix $0$ denotes variables evaluated 
 at the beginning of the time 
step, and other variables are calculated at the end of the time step.  
 $T$, $S$, $P$, $\rho$, $L$ denote the temperature, the entropy, 
 the pressure, the 
density and the luminosity respectively,
$q$ denotes the rate of nuclear energy production (minus the energy
carried by neutrinos),  $v$ is the velocity of matter, and $x_a$ and $R_a$
are the chemical composition and the reaction rate of the chemical
isotope $a$, respectively.
 While quasi-hydrostatic evolution holds, the left side of equation (3) is 
equated to zero.
The fully dynamical version can be used during fast
evolutionary phases, which are observed whenever $(v - v^0)/\delta t $ 
exceeds a few percent of $Gm/r^2$  anywhere.  
 Nuclear reactions are calculated for five elements ($H,He,
C+N, O, Ne$). 
 The equations are implicit, and are solved by 
iterations, using the full Heney code.  

Convection is calculated by the mixing length prescription: 
\begin{equation}
L_{conv} = 4\pi r^2 \rho C_p v_c l_p \Delta T ,
\end{equation}
where 
\begin{equation}
\Delta T = \left( \vert{dT\over dr}\vert_{star} - 
\vert {dT\over dr}\vert_{ad} \right) ,  
\end{equation}
and 
\begin{equation}
v_c^2 = l_p^3 \Delta T . 
\end{equation}
Here $v_c$ is the convective velocity, $l_p$ is the pressure scale height,
and $C_p$ is the heat capacity per unit mass at constant pressure.

The convection velocity is given in units of the local isothermal
sound velocity $c_s= (kT/\mu m_H)^{1/2}$.
 From these equations we derive the the ratio of convective to
thermal pressure
\begin{equation}
\frac {P_{\rm conv}}{P_{th}} =
\frac {v_c^2}{c_s^2} =
\left( \frac {g l_p}{2 T C_p} \right)^{2/3} 
\frac {F_{\rm conv}^{2/3}} {c_s^2}
\rho^{-2/3} .
\end{equation}
We limited the value of $v_c/c_s$ to unity, and in the zone in which 
this value equals unity, the convective ram pressure is actually of 
the same magnitude as the thermal pressure.  
The atmosphere of the star was calculated by using Eddington 
approximation for grey atmosphere, hence we do not discuss details 
of the photosphere structure, and the values presented in
the figures below are not accurate close to the stellar surface.

{{{  Several notes should be made regarding the numerical calculation.
 First, we are interested mainly in the relative changes in the
properties of the evolving envelope.
 Therefore, changing numerical values of physical variables that influence
all evolutionary models in the same sense, e.g., the mixing length,
will not affect our results.
 We use values that were found by us in previous works to give the
best results.
 Second,  the numerical code does not include dust formation in the
atmosphere, and we do not have dust opacity.
{{{{ This code was developed using opacities from Alexander 1975.}}}}
 New opacity tables {{{{ (e.g., Alexander, Rypma \& Johnson 1983) }}}}
were incorporated as they became available
since this code was first used by us (Harpaz \& Kovetz 1981), when they
were found to be significantly different from former tables.
 Third, we do not model a full evolutionary track, but only calculate
the structure of the envelope at five evolutionary points.
 Only in moving from the first to second point we have also evolved the
core,
{{{{ for an average mass loss rate of $10^{-6} M_\odot \yr ^{-1}$ }}}} .
 In the last 4 evolutionary points the core mass was held constant
(see below).
 Hence, we do not have a mass loss rate formula, but mechanically remove
mass from the envelope until the desired envelope mass is achieved. 
}}}

 Using this numerical code, we evolve a stellar model along the
the AGB and beyond. 
In Figures $1-5$ we present some of the envelope properties versus the
radius, at five points along the evolution.
 The quantities that are plotted
are the temperature $T$ (in Kelvin), density $\rho$ ($\g \cm^{-3}$), 
the mass $m$ ($M_\odot$), the entropy $S$ (in relative units), 
the convection velocity $v_c$ (in units of the local sound speed $c_s$),
the thermal pressure $P$ (dyne$~\cm^{-2}$), the moment of inertia $I$ 
($M_\odot R_\odot^2$), and the pressure scale height $l_p$ ($R_\odot$).  

The envelope masses at the five evolutionary points are
$M_{\rm env}= 0.5$, $0.3$, $0.1$, 0.033,
and $0.015 M_\odot$, respectively.  
 At an early time, when the mass is $1.1 M_\odot$, the
core mass is $0.58 M_\odot$.
 We assume that when the envelope mass becomes $0.3 M_\odot$ the superwind
started, and the core mass does not evolve much further.  
In order to isolate the influence of the envelope parameters during this 
evolutionary phase, we kept the core constant by switching off the 
chemical evolution.  
 The core mass was kept at $0.6 M_\odot$.

% ======================================================================
\section{DISCUSSIONS}
% ======================================================================
% ======================================================================
\subsection{The Density Profile}
% ======================================================================

 The most striking changes in the envelope due to the mass loss
are the decrease in the density below the photosphere
and the changes in the density and entropy profiles
{{{  (figs. 6 and 7, respectively). }}}
 To emphasize this behavior, we present the density profile of the
five models in Figure 6. 
 This can be understood as follows.
  The photospheric pressure $P_p$ and density $\rho_p$ are determined
by the stellar effective temperature $T_p$, luminosity, and photospheric
opacity $\kappa$, and are given by (e.g.,  Kippenhahn \& Weigert 1990, $\S 10.2$)
\begin{equation}
P_p= 
\frac {2}{3} 
\frac {G M}{R^2} 
\frac {1} {\kappa},
\end{equation}
and 
\begin{equation}
\rho_p = 
\frac {2}{3} 
\frac {G M \mu m_H}{k_B} 
\frac {1} {R^2 \kappa T_p},
\end{equation}
where $M$ is the stellar mass, $R$ is the photospheric radius, $k_B$ is
the Boltzmann constant, and $\mu m_H$ is the mean mass per particle.
 In deriving these expressions we have used the definition of the
photosphere as the place where $\kappa \l \rho_p =2/3$,
where $l$ is the density scale height. 
  At the level of accuracy of these expressions, we can take the
pressure and density scale height at the photosphere to be equal (we do 
not consider here the density inversion region below the photosphere).

 In the range of the effective temperatures $2800 < T_p < 3600$ and the
typical photospheric density of AGB stars, we find from Table 6 of
Alexander \& Ferguson (1994) that we can take the photospheric opacity
to be (for a solar composition)
$\kappa \simeq 4 \times 10^{-4} (T/3,000)^q \cm^2 \g^{-1}$, with $q \sim 4$.
 Substituting typical values for an upper AGB star, we find the ratio of
the photospheric density to the average envelope density
$\rho_a = 3 M_{\rm env} /(4 \pi R^3)$ to be 
\begin{equation}
\frac {\rho_p}{\rho_a} \simeq 0.25
%
%  0.66666667  *  G  M  \mu M_H   *4 \pi R *   10 
%----------------------------------------------
%  3 \times 10^{-4} * 3,000 * k     * 3
%
\left( \frac{R}{300 R_\odot} \right)
\left( \frac{M}{10 M_{\rm env}} \right)
\left( \frac{T_p}{3,000 \K} \right)^{-q-1} .
\end{equation}
  As the star evolves along the AGB, the three factors of equation (12)
contribute to the increase in the ratio of the photospheric density to
the average density, with the envelope mass being the most influential. 
 Therefore, the increase in this ratio is quite fast on the upper AGB, as
the envelope mass decreases.
  This ratio continues to increase even as the envelope starts to contract,
when its mass is $M_{\rm env} \sim 0.1 M_\odot$.
 We find here (see also fig. 3 by Soker 1992) that in the range
$5 \times 10^{-3} \lesssim (M_{\rm env}/M_\odot) \lesssim 0.1$
the envelope radius goes as $R \propto M_{\rm env}^{0.2}$.
 For a constant luminosity post-AGB star, the temperature goes as
$R^{-1/2}$, and with $q=4$ for the opacity dependence, we find
\begin{equation}
\frac {\rho_p}{\rho_a} \simeq
\left( \frac {M_{\rm env}}{0.1 M_\odot} \right)^{-0.3}
\left( \frac {\rho_p}{\rho_a} \right)_{M_{\rm env} = 0.1} .
\end{equation}
 This predicts a very shallow density profile for post-AGB stars with 
$M_{\rm env} \gtrsim 5 \times 10^{-3} M_\odot$, as is indeed found by 
Soker (1992; see his fig. 1) and the previous section here. 
 During this stage the radius is still large, $R \gtrsim 150 R_\odot$,
and the temperature low enough for dust to form quite easily. 
 Therefore, the deviation from spherical mass loss, if facilitated by the
shallower density profile, may increases during the early post-AGB phase.

% ======================================================================
\subsection{The Angular Momentum Problem}
% ======================================================================

The changes in the envelope's properties should more than
compensate for the fast decrease in the angular velocity of the star.  
 The angular velocity decreases very fast due to the intense mass loss
rate. 
 For an envelope density profile of $\rho \propto r^{-2}$, the
decrease goes as (for a constant radius on the upper AGB and a solid body
rotation in the envelope)
$\omega \propto M^2_{\rm env}$ (Harpaz \& Soker 1994). 
 Since the density profile is steeper than $\rho \propto r^{-2}$ during
most of the early AGB phase, the decrease in the angular velocity will be
even faster.

 Let us refer to the dynamo activity required for the formation of
cool magnetic spots (Soker 1998).
At this point we cannot predict the angular velocity which is required to
operate an efficient dynamo in AGB stars.
However, based on the strong convective motion, we speculate that the
required angular velocity is very low.
 Soker (1998), based on the work of Soker \& Harpaz (1992),
{{{  crudely }}} estimates the required equatorial surface angular
velocity to be $\omega \gtrsim 10^{-4} \omega_{\rm Kep}$,
where $\omega_{\rm Kep}$ is the Keplerian velocity on the equator.
 For angular velocity of $\omega \gtrsim 10^{-2} \omega_{\rm Kep}$,
a massive companion is required to spin up the envelope, and other
effects become important.
 Therefore, the mechanism of cool magnetic spots, though it can be very
effective for fast rotations, was introduced in order to explain
axisymmetrical mass loss from slow AGB rotators, i.e.,
$\omega \lesssim 10^{-2} \omega_{\rm Kep}$.
 An angular velocity of $\omega \sim 10^{-4} \omega_{\rm Kep}$ can be
attained even by fast rotating main sequence stars.
 However, due to the angular momentum loss mentioned above, in order for
the dynamo to stay effective, the AGB star should be spun-up by a
companion, and/or the dynamo {{{  must be }}} effective even for
$\omega \sim 10^{-5} \omega_{\rm Kep}$.
 As pointed out by Soker (1998), for the envelope spin-up,
if it occurs on the upper AGB, a planet companion of mass
$\sim 0.1 M_{\rm Jupiter}$ is sufficient.
 As we mention below, born-again AGB stars may hint that the
mechanism is efficient even for
 $\omega \sim 0.3 \times 10^{-4} \omega_{\rm Kep}$.

  As pointed out by Soker (1998) and Soker \& Clayton (1999),
possible support for the influence of the low density in the envelope,
and the effectiveness of the mechanism for axisymmetrical mass
loss, comes from the PN A30.
 This PN has a large and almost spherical halo, with optically bright 
hydrogen-deficient blobs in the inner region, which are arranged in
a more or less axisymmetrical shape.
 The blobs are thought to result from a late helium shell flash
(i.e., a born-again AGB  star).
%(REFERENCE--from Chu??)
 After a late helium flash, the star expands to a radius of
$R \gtrsim 100 R_\odot$, and since it has a very low envelope mass, the
density profile will be very shallow.
 This may explain the axisymmetrical knots of A30, despite its
almost spherical halo.
Soker \& Clayton (1999), by using the dust formation above cool magnetic
spots, point to a possible connection between the mass loss behavior
of R Coronae Borealis stars, which are thought to be born-again stars
and AGB stars.

Heber, Napiwotzki, \& Reid (1997) find that single WDs rotate very
slowly, $v_{\rm {rot}} <  50 \km \s^{-1}$.
 A central star of a PN which rotates at a velocity of $10 \kms$, when
expanding as a born-again AGB star to $\sim 100 R_\odot$, will
rotate at $\sim 10^{-3} \km \s^{-1} \simeq 0.3 \times 10^{-4}$$~v_{\rm Kep}$,
where $v_{\rm Kep}$ is the Keplerian rotation velocity.
 A shallow density profile with some dynamo activity may result in dust
formation (as in R Coronae Borealis stars which have similar radii), and
axisymmetrical mass loss.

 Soker (1998) extensively discusses the advantage of the cool magnetic
spots mechanism over models which require fast AGB rotation (his section 2).
Despite this, in a recent paper Garcia-Segura {\it et al.} (1999) propose
a scenario which they claim can operate for {\it single} AGB stars.
 They argue for an equatorial rotation velocity of
$>1 \kms$ for an AGB star of $150 R_\odot$.
  We think their model can operate only if a stellar companion
of mass $M_2 \gtrsim 0.1 M_\odot$ spins-up the envelope.
  But then other effects of the companion 
(e.g., Livio 1997; Soker 1997; Mastrodemos \& Morris 1998; 1999), 
become more significant. 
 More specifically, we think their model is wrong for the following reasons.
(1) Their mechanism to spin-up the envelope on the upper AGB is a transfer
of a large amount of angular momentum from the core to the envelope of
mass $\sim 0.1 M_\odot$.
 However, a coupling between the core and the envelope occurs at early
stages of the evolution, i.e., first and second dredge-up, on the RGB and
early AGB, respectively.
(2) Many bipolar PNs have equatorial mass concentration with a mass of
$\gg 0.1 M_\odot$, for which their mechanism of spin-up is not efficient.
(3) Even for an envelope mass of $0.1 M_\odot$, the required angular
momentum of the core means that it rotates, before its coupling to the
envelope, at an angular velocity of $\sim 0.3$ times the Keplerian velocity
at the core's surface.
 This seems to be too fast.
 Balbus \& Hawley (1994) maintain that the powerful weak-field MHD instability
is likely to force a solid body rotation even in the radiative zone
of stars.
  We think that this MHD instability is likely to transport most of the
core's angular momentum to the envelope, even if the region is stable
to the Hoiland criterion.
(4) For the formation of bipolar PNs their model requires the rotation
velocity to be $\sim 6.9 \kms$ (their model D), for which they get an
equatorial to polar density ratio of 112.
By the time the rotation velocity is $6.36 \kms$ (their model C),
the density ratio decreases to 9.
 However, as we discussed above, a decrease of the envelope mass by only
$\sim 4 \%$ will bring their model D to model C.
A loss of $\sim 30 \%$ of the envelope will bring them to model B,
a rotation velocity of $3.5 \kms$ and a density contrast of  $<2$.
(5) To support their model, they cite the similarity between the structure
of the Hourglass PN (also named MyCn 18 and PN G 307.5-04.9) and the
nebula around $\eta$ Carinae.
 However, it seems that the central star of $\eta$ Carinae
has a binary companion at an orbital separation of $\sim 15 \AU$
(e.g., Lamers {\it et al.} 1998).
(6) The problems with the magnetic activity during the late post-AGB
phase were listed by Soker (1998, his section 2.2),
and we will not repeat the arguments here.
 Basically, this mechanism requires substantial spin-up by a massive
binary companion.
 We think that each of the first four points above is strong enough to
make their model questionable.
 
% ======================================================================
\subsection{Implications for Magnetic Cool Spots}
% ======================================================================

 Our knowledge of magnetic cool spots comes mainly from the sun, for which
there is  a huge amount of theoretical work (e.g., Priest 1987),
although there is no complete acceptable theory yet for the formation
and evolution of spots. 
 However, some basic ingredients seem to be common to all models. 
\newline
1) First, there is a need for dynamo activity to amplify the magnetic field,
and replace the magnetic flux which reaches the solar surface and escapes.
 In the previous subsection we presented some indications that
the mechanism for axisymmetrical mass loss may operate for very slow
rotation $\omega \gtrsim 3 \times 10^{-5} \omega_{\rm Kep}$. 
 We suggest that due to the strong convective motion, i.e., high 
convective pressure, which results from the steep entropy gradient,
dynamo activity in AGB stars is possible even when the
rotation is very slow. 
No theory for dynamo activity exists which allows us to calculate the 
magnetic activity, and therefore at present we cannot expand on this point. 
\newline
2) Another basic ingredient for the formation of cool magnetic spots
is the process by which the motion of convection cells concentrates
magnetic flux to form a strong vertical magnetic field, which then
suppresses the vertical convective heat transport, hence leading to
a cool spot.
 We suggest that the high convective pressure makes this process
more efficient.
\newline
3) In addition to the external convective pressure that concentrates the
magnetic field, there is another stage in the amplification of the
magnetic field inside the tube.
In this stage material cools, because of the reduced heat transfer, and
sinks inside the tube (Meyer {\it et al.} 1974; Priest 1987. $\S 8.6.1$).
 The steeper entropy profile (fig. 7) makes the envelope, inside and
outside the tube, more prone to any convective motion.
Hence, the sinking of cool gas mechanism to enhance the magnetic field
{{{  may }}} become more efficient.
\newline
4) For the formation of a stable spot the flux tube should be vertical.
  The sensitivity of the envelope to convective motion (i.e. to the
sinking and rising motion of blobs), because of the steep entropy
gradient, may cause rising flux tubes to become vertical more frequently.

 That the convective pressure increases can be seen from simple
considerations that led to equation (9).
 From all factors in that equation, only the density changes fast
during the intense mass loss rate on the upper AGB.
 The density decreases, and hence the convective to thermal pressure
ratio increases.
 As can be seen from figures 1-5, the ratio of convective to isothermal
sound speed becomes equal to unity (i.e., $v_c/c_s=1$) deeper and deeper
in the envelope.
 This, we suggest, results in strong magnetic activity in the outer
envelope. 

{{{  The solar magnetic field may be used as a hint on the expected
behavior of AGB stars magnetic field, despite the large
differences between the types of stars (Soker \& Clayton 1999).
 This is implied from our assumption that dynamo mechanism amplifies
the AGB magnetic field, as is the case for the solar magnetic field.
 Most relevant to the present work are the 4 basic ingredients listed
above, and the expected concentration of cool magnetic spots toward
the equator of the star. }}}

 Finally, two other interesting properties should be noted.
First, the location of the inner boundary of the convective region
moves outward.
 For the envelope masses of $0.5$, $0.3$, $0.1$, $0.033$, and 
$0.015 M_\odot$, the inner boundaries are at $0.3$, $0.4$, $1$, $4$,
and $6 R_\odot$, respectively. 
Second, the photospheres of cool magnetic spots in AGB stars are
{\it above} the rest of the photosphere
(Soker \& Clayton 1999), contrary to the sun where cool spots
are several hundred kilometers deeper than their surroundings.
This complication should be included in future detailed calculations
of the structure of magnetic cool spots in AGB stars.

% ======================================================================
\section{SUMMARY}
% ======================================================================

 In the present paper we examined the possibility that the transition
from a spherical mass loss to an axisymmetrical mass loss on the
upper AGB is caused by changes in some of the envelope's properties.
 The transition to axisymmetrical mass loss is inferred from the structure
of many elliptical planetary nebulae.
 Our main results can be summarized us follows.
\newline
 (1) As the envelope mass decreases on the upper AGB
and early post-AGB phases, the density quickly decreases and its
profile becomes very shallow.
 At the same time the entropy profile becomes very steep.
 This suggests that the transition to axisymmetrical mass loss
{{{  which is inferred from observations, }}}
may be related to these properties, unless it is due to 
a late interaction with a stellar or a planet companion.
\newline
(2) We {{{  qualitatively }}} discussed a few processes by which the
changes in the {{{  density and entropy profiles }}}
may lead to an enhanced formation of magnetic cool spots,
{{{  which we assumed to be concentrated }}} near the equatorial plane.
{{{  We suggest that }}}
this in turn will lead to an enhanced formation of dust and a higher mass
loss rate near the equatorial plane.
{{{  We also claim that }}}
such magnetic activity requires only slow rotation, as the magnetic
field is not globally dynamically important.

{\bf ACKNOWLEDGMENTS:} 

 We would like to thank the referee, S. H\"ofner, for detailed
and helpful comments.
 This research was supported in part by a grant from the University
of Haifa and a grant from the Israel Science Foundation.

\newpage 

{\centerline {\bf FIGURE CAPTIONS}}

\no {\bf Figure 1:} 
 The envelope structure of an AGB star model with a total mass of
 $1.1 M_\odot$, a core mass of $0.58 M_\odot$,
and a luminosity of $8,400 L_\odot$. 
  The quantities that are plotted versus the radius are: 
temperature $T$, density $\rho$, 
the mass $m$ ($M_\odot$), the entropy $S$, 
the convection velocity $v_c$ (in units of the local sound speed),
the thermal pressure $P$, the moment of inertia $I$ 
($M_\odot R_\odot^2$), and the pressure scale height $L_p$ ($R_\odot$).  
$T$, $\rho$, and $P$ are in cgs units, and $S$ is in relative units.
 Note that we treat the region near the photosphere using the 
Eddington approximation of grey atmosphere, and therefore 
the values of the density, pressure, and temperature 
very close to the surface (photosphere) are not accurate.

\noindent {\bf Figure 2:} 
 Like figure 1, but at a later time when the total mass is $0.9 M_\odot$
and the core mass is $0.6 M_\odot$.
 The stellar luminosity is $8,700 L_\odot$.

\no {\bf Figure 3:} 
 Like figure 2 {{{  (the same core mass and luminosity), }}}
but at a later time when the total mass is $0.7 M_\odot$.

\no {\bf Figure 4:}
 Like figure 2, but at a later time when the total mass is $0.63 M_\odot$.

\no {\bf Figure 5:}
 Like figure 2, but at a later time when the total mass is $0.615 M_\odot$.

\no {\bf Figure 6:}
 The density profile of the five models presented in figs. 1-5.
 The thick lines represent the model during the contraction phase,
while the thin solid line is the model presented on fig. 1
(total mass of $1.1 M_\odot$).
 The masses of the models are: $0.9 M_\odot$ (dot, dot, dot-dash line),
$0.7 M_\odot$ (dot-dash line), $0.633 M_\odot$ (dotted line),
and $0.615 M_\odot$ (solid line).

\no {\bf Figure 7:}
{{{   The entropy profile of the five models presented in figs. 1-5.
 The lines represent the same models as in figure 6. }}}

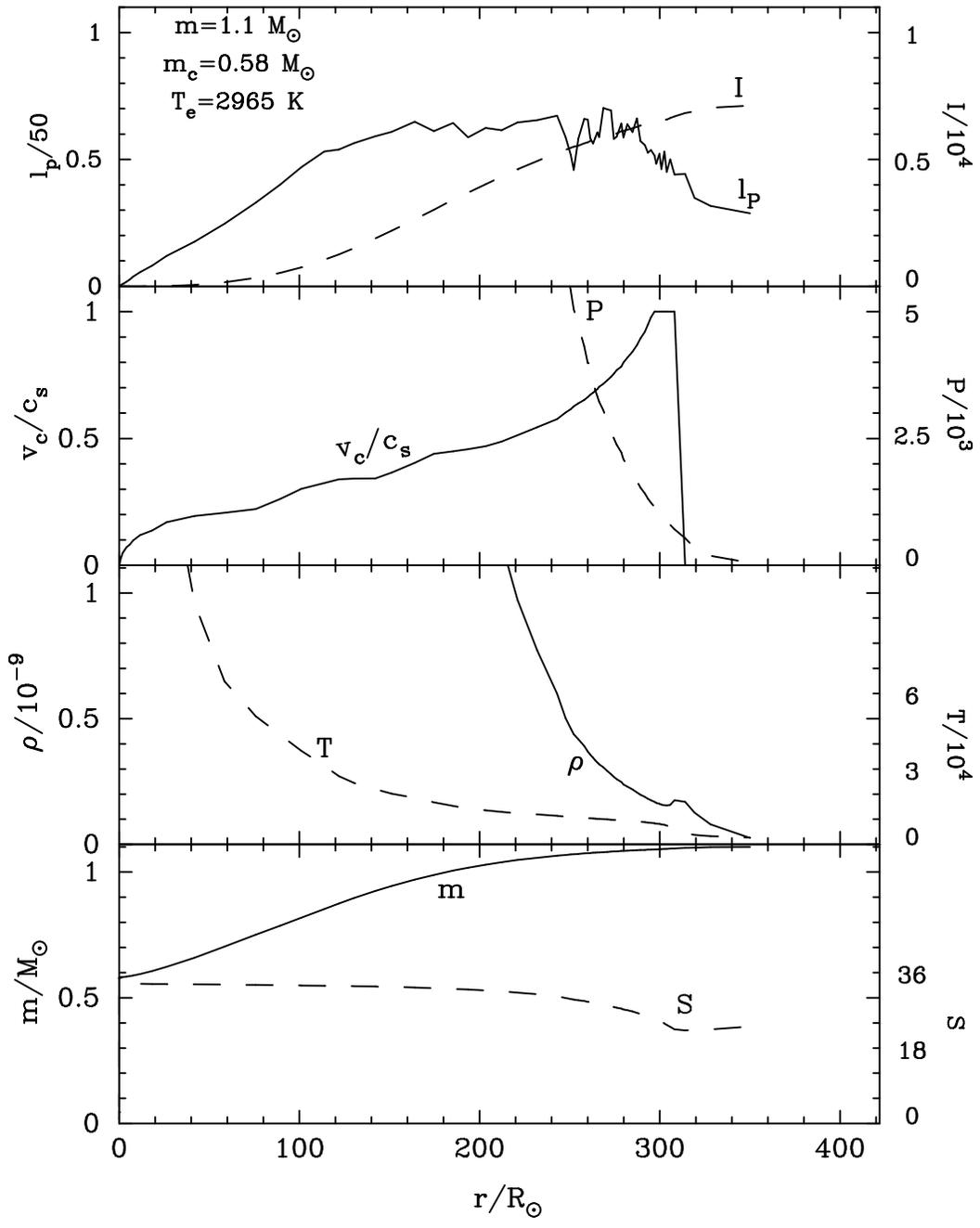
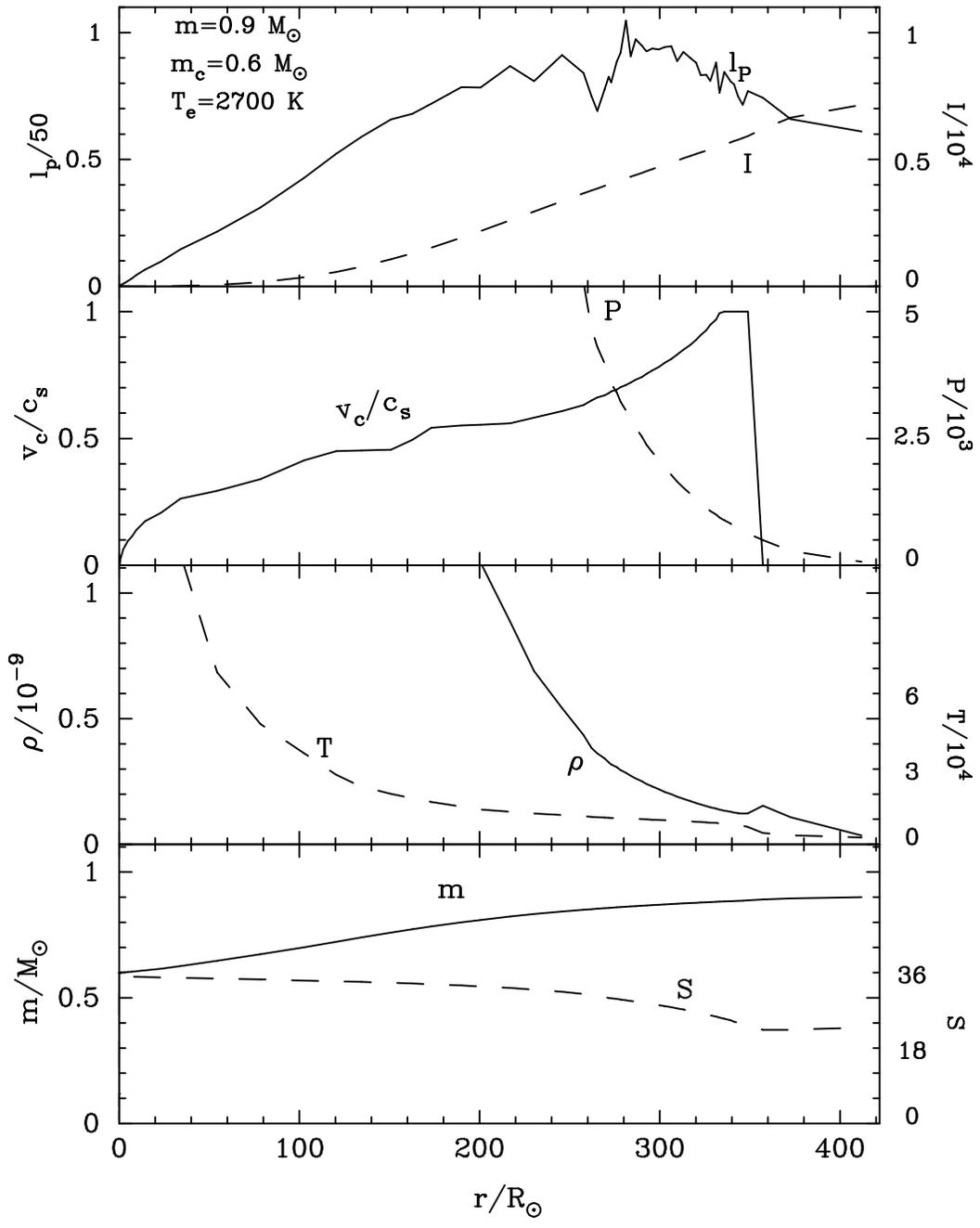
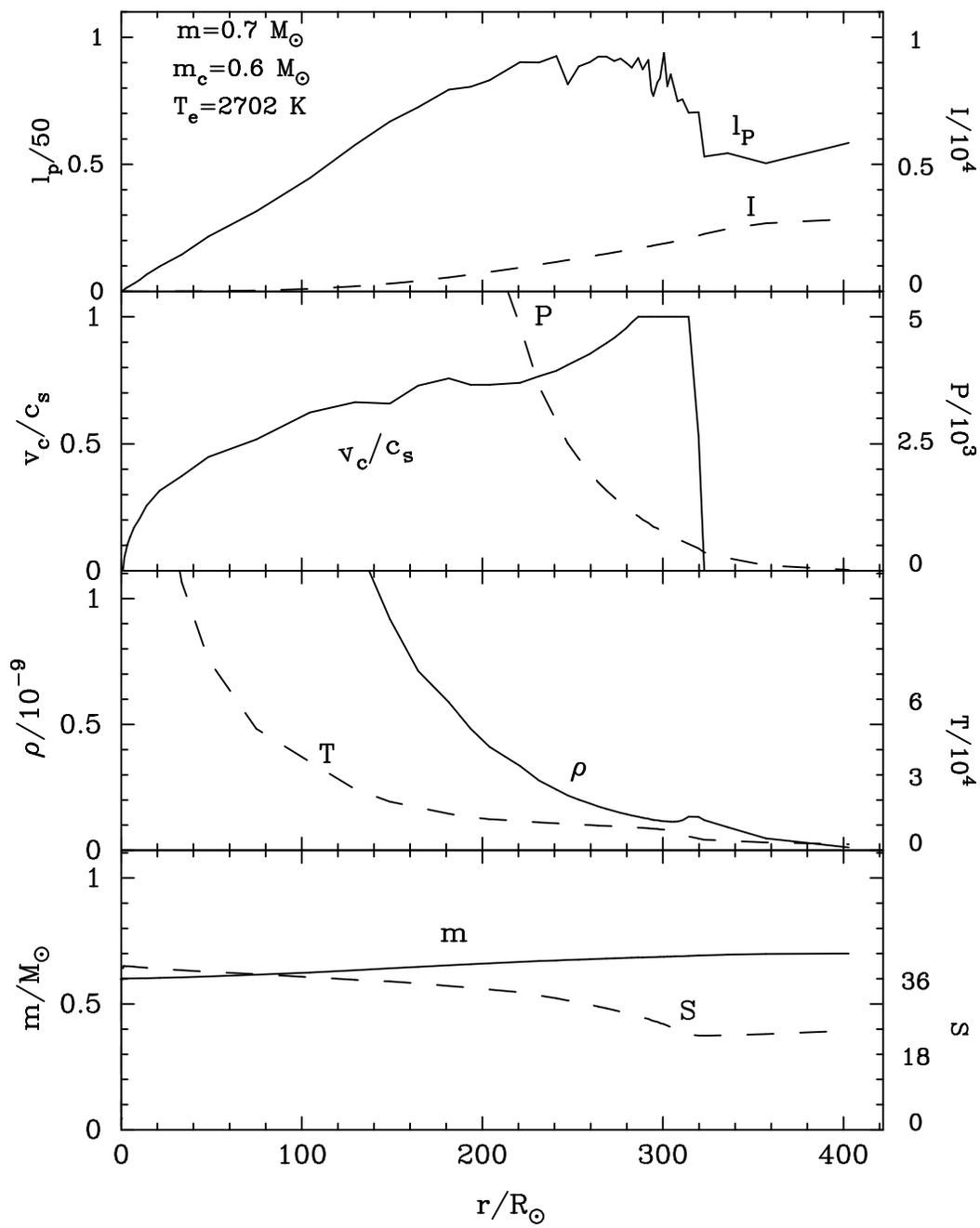
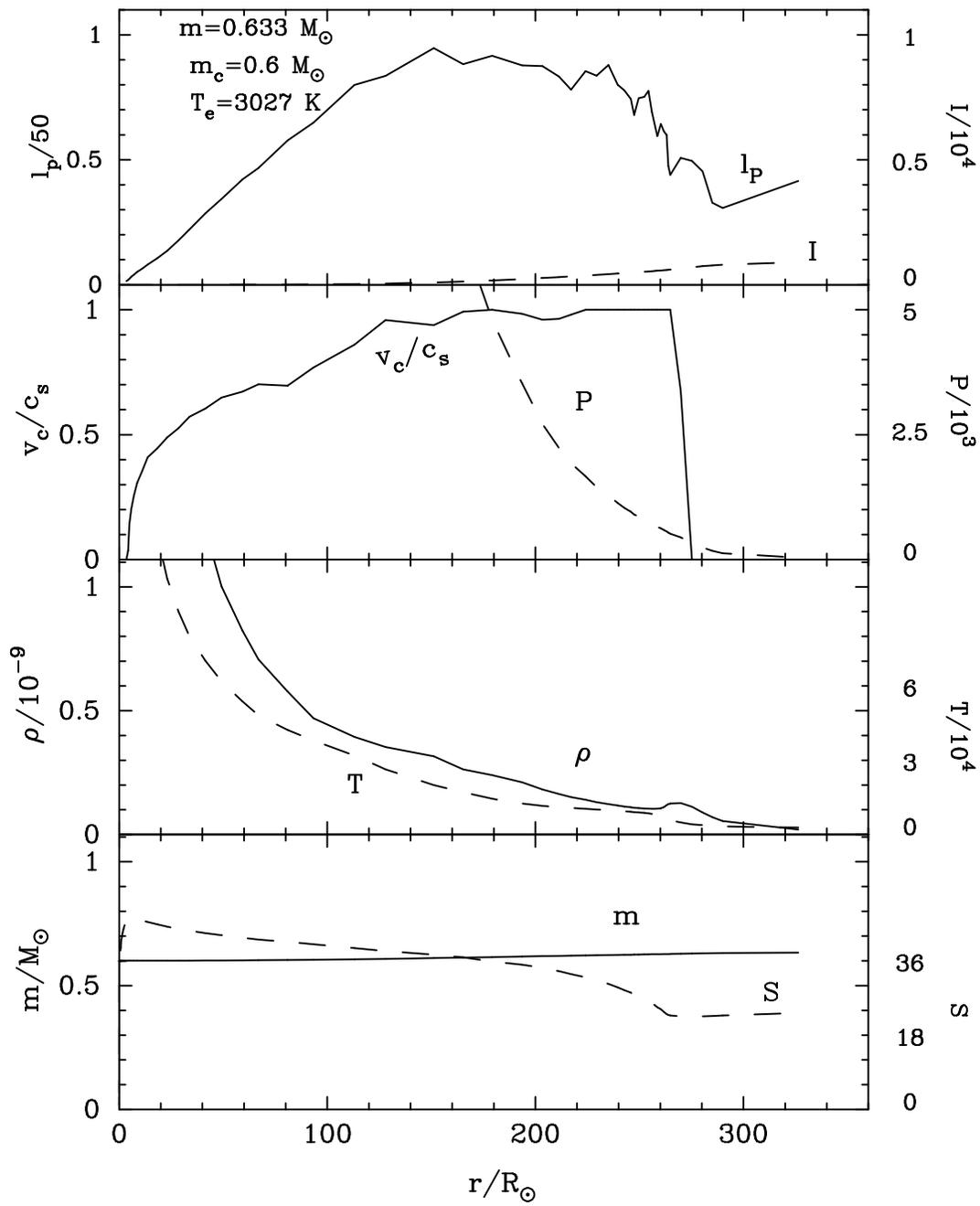
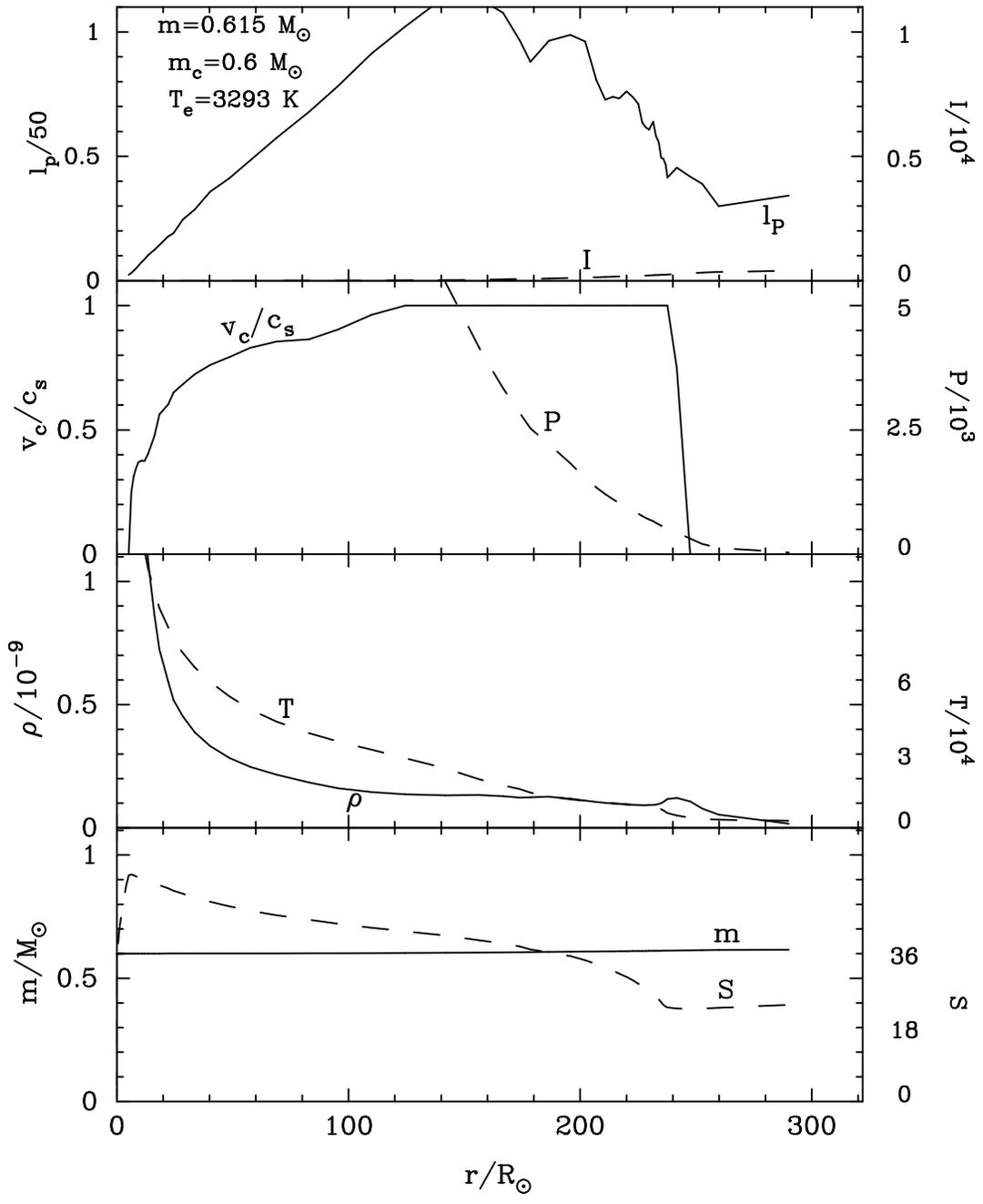
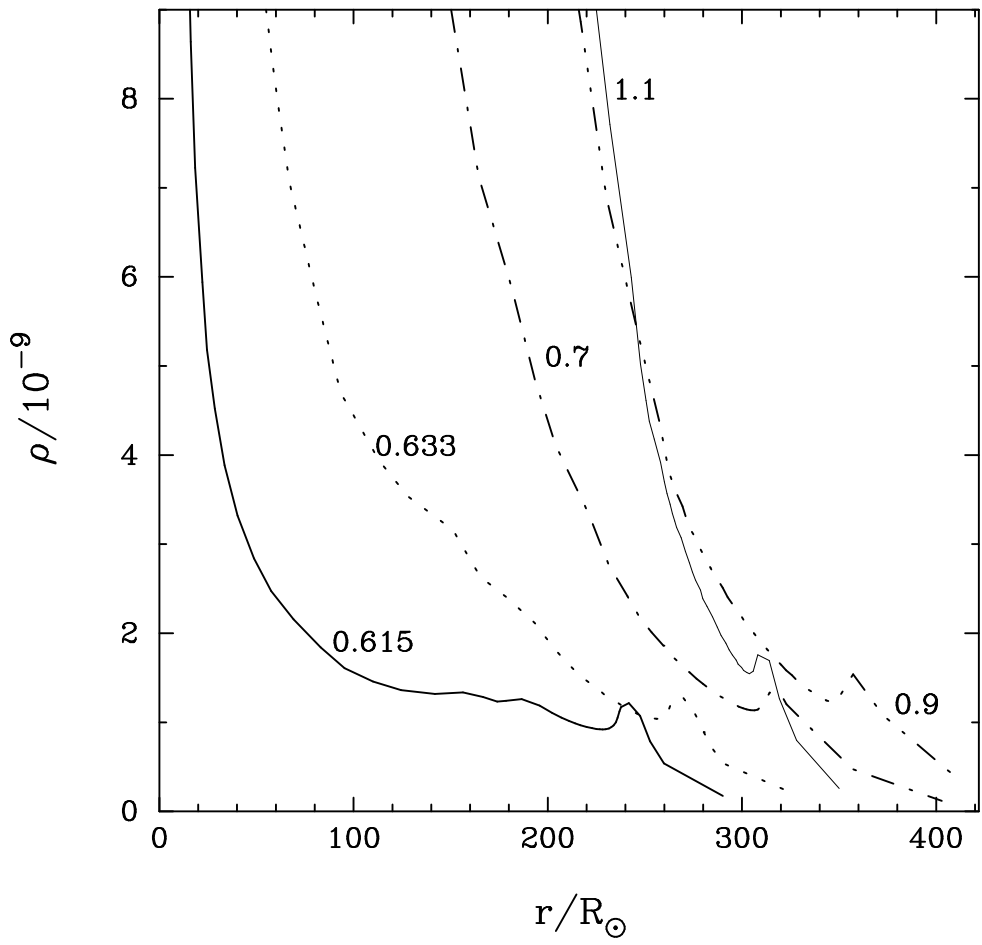
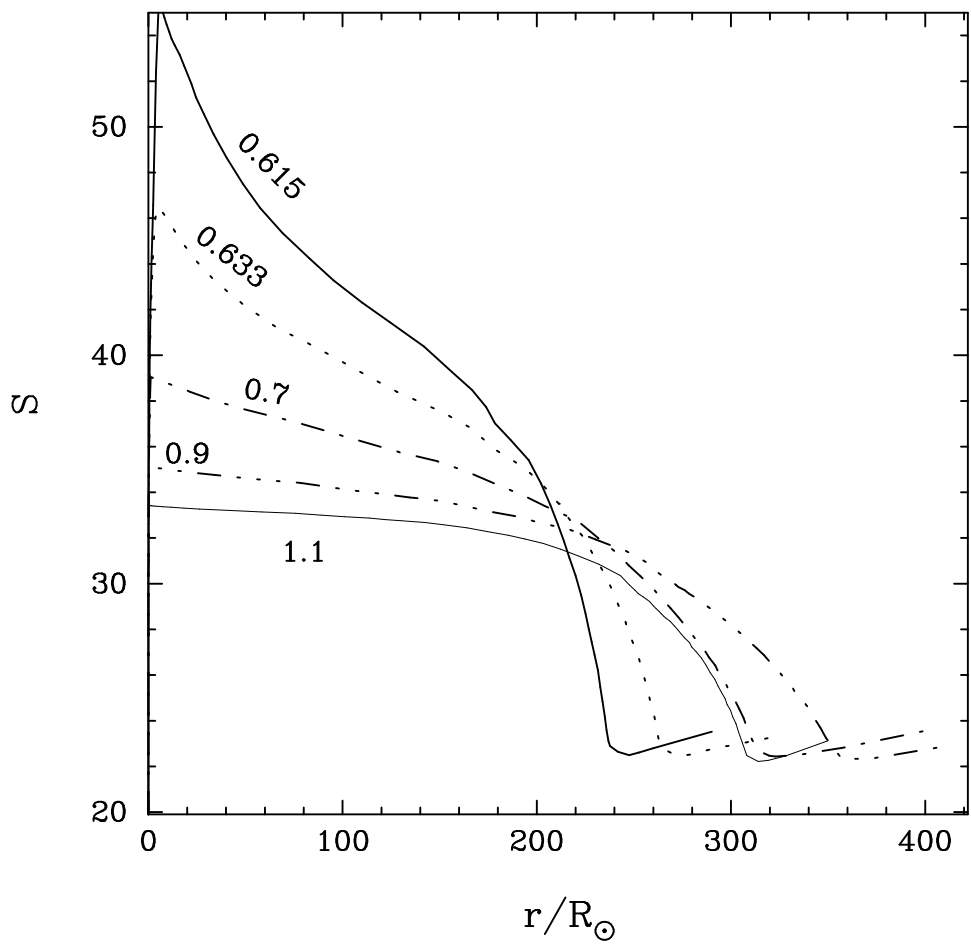
\end{document}